\documentclass[conference]{IEEEtran}
\IEEEoverridecommandlockouts
\usepackage{cite}
\usepackage{amsmath,amssymb,amsfonts}
\usepackage{algorithmic}
\usepackage{graphicx}
\usepackage{textcomp}
\usepackage{xcolor}
\usepackage{subfigure}
\usepackage{multirow}
\def\BibTeX{{\rm B\kern-.05em{\sc i\kern-.025em b}\kern-.08em
    T\kern-.1667em\lower.7ex\hbox{E}\kern-.125emX}}
\begin{document}

\DeclareRobustCommand*{\IEEEauthorrefmark}[1]{%
    \raisebox{0pt}[0pt][0pt]{\textsuperscript{\footnotesize\ensuremath{#1}}}}

\title{Multi-Frequency Wireless Channel Measurements and Characteristics Analysis in Indoor Corridor Scenarios}

\author{\IEEEauthorblockN{Zihao~Zhou\IEEEauthorrefmark{1,2}, 
Li~Zhang\IEEEauthorrefmark{1,2}, 
Xinyue~Chen\IEEEauthorrefmark{1,2},  
Cheng-Xiang~Wang\IEEEauthorrefmark{1,2*},  
and Jie~Huang\IEEEauthorrefmark{1,2} }
 \IEEEauthorblockA{$^1$National Mobile Communication Research Labratory, School of Information Science and Engineering,\\ 
 Southeast University, Nanjing 210096, China \\ 
 $^2$Purple Mountain Labratories, Nanjing 211111, China\\
 $^*$Corresponding Author: Cheng-Xiang Wang\\
 Email: \{zhouzh, li-zhang, chenxinyue\_2019, chxwang, j\_huang\}@seu.edu.cn}}

\maketitle

\begin{abstract}
In this paper, we conduct wireless channel measurements in indoor corridor scenarios at 2.4, 5 and 6 GHz bands with bandwidth of 320 MHz. 
The measurement results of channel characteristics at different frequency bands such as average power delay profile (APDP), path loss (PL), delay spread (DS), and Ricean K factor (KF) are presented and analyzed.
It is found that the PL exponent (PLE) and PL offset $\beta$ in the floating-intercept (FI) model tend to increase with the increase of frequency. 
The DS and KF values of the three frequency bands in line of sight (LOS) scenario are basically the same. 
These results are significant for the design of communication systems. 
\end{abstract}


\section{Introduction}
With the development of communication technology, sub-6 GHz spectrum resources are becoming increasingly scarce. Therefore, as a technology to improve spectrum efficiency, multi-frequency cooperation has become a hot issue \cite{a1}. 
Various technologies used in indoor short-range wireless communication usually work in multi-frequency bands, such as Zigbee working at 868 MHz, 915 MHz, and 2.4 GHz, wireless local area networks (WLAN) working at 2.4 GHz, 5 GHz and possibly 6 GHz in the future. 
The existence of these devices inspires the possibility of multi-frequency cooperation technology. 
The statistical characteristics of multi-frequency channels provide important guidance for channel model performance evaluation \cite{ch1} and communication system design\cite{new1,new2,new3}.\par
As for the research of multi-frequency channel characteristics, most of the existing channel measurements compared the channel characteristics difference bentween sub-6 GHz and millimeter wave.
In \cite{MF1}, the channel measurement was conducted under 2-4, 9-11, and 27-29 GHz. The measurement results show that the PL offset in the FI model 
increases with the increasing frequency. In \cite{MF2}, the authors studied the difference between DS and direction spread
at 5.8, 14.8, and 58.7 GHz, and no obvious frequency dependence of DS and direction spread was found. Channel measurements at 3, 10, and 28 GHz were
carried out in \cite{MF3}, \cite{MF4}, the authors found that the diffuse components dominate at 3 GHz, however the specular component takes up 
more power at 10 and 28 ~GHz. There are few measurements of multi-frequency correlation specially under sub-6 GHz. 
The DS and coherence bandwidth of 2, 4, and 6 GHz with 200 MHz bandwidth were measured in \cite{a6}, but this measurement  has only one sample in LOS and NLOS scenarios, respectively.
In \cite{a7}, the channel spatial characterization at 2.45 and 6.05 GHz in the classroom scenario was investigated. The spatial correlation at 6.05 GHz is smaller than 2.45 GHz in the case of same antenna spacing. 
In \cite{MF5},channel measurement was conducted in indoor scenario at 2.45 and 6.05 GHz bands, the authors mainly analyzed the difference in PL and DS between these two frequency bands.\\
\indent In the sub-6 GHz channel measurements mentioned above, the channel  bandwidth was limited within 200 MHz in \cite{a6}--\cite{MF5}, which may be less
than that of some future wireless communication systems. In addition, the analysis of channel characteristics in these papers is not comprehensive enough. 
In \cite{a6}, only one position point was measured in LOS and NLOS scenarios, respectively. In \cite{MF5,a7}, only two frequency bands were measured.
In this paper, channel measurement at 2.4, 5, and 6 GHz with bandwidth of 320 MHz is conducted in 
indoor corridor environment. 
A  total  of  69  locations are measured in LOS and non LOS (NLOS) scenarios, and then the channel statistical characteristics, such as 
APDP, PL, DS and KF are analyzed comprehensively.\\
\indent The remainder of this paper is organized as follows. Section~II introduces the channel sounder and measurement environment. Section III introduces the 
data processing method. Measurement results and detailed channel characteristics analysis are presented in Section IV. Finally, conclusions are drawn in Section IV.

\section{Measurement Campaign}

\subsection{Measurement System Setup}

\begin{figure}[!t]
    \centerline{\includegraphics[width=8cm]{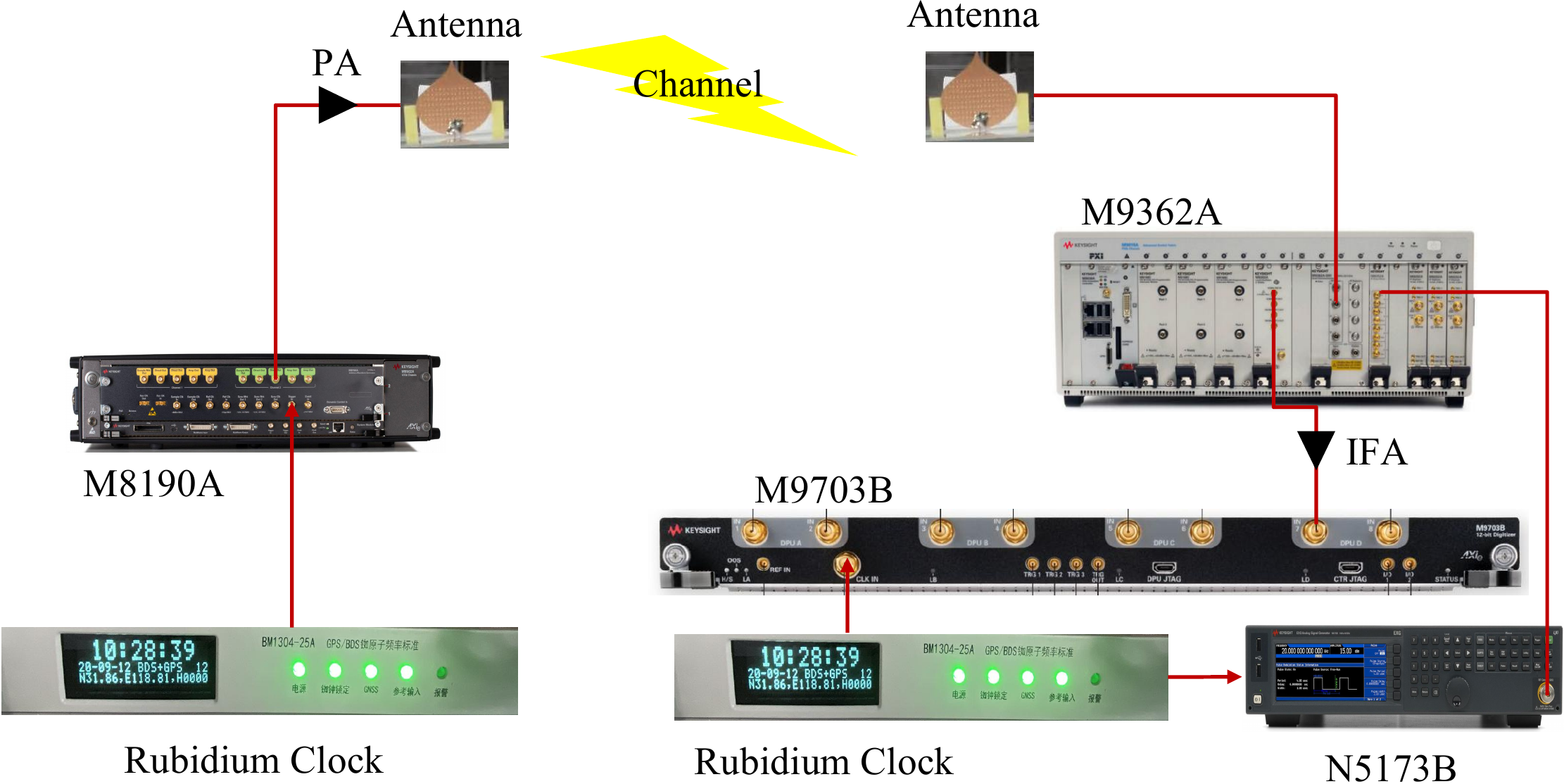}}
    \caption{Diagram of Keysight time domain channel sounder.}
    \label{diagram}
\end{figure}

This multi-frequency channel measurement campaign was conducted by Keysight time domain channel sounder illustrated in Fig. \ref{diagram}. 
This system can fully support sub-6 GHz indoor channel measurement with signal 
bandwidth of 320 MHz. Table \ref{table1} lists a detailed configuration and parameters of this measurement.

\indent As shown in Fig. \ref{diagram}, the transmitter (Tx) side consists of a M8190A arbitrary waveform generator, a high-precision GPS Rubidium clock, 
and a power amplifier (PA). The receiver (Rx) side consists of a M9362A PXIe down converter, a M9703B 12-bit broadband digitizer, a N5193B vector signal generator, a 
high-precision GPS Rubidium clock, and an intermediate frequency amplifier (IFA). Omni-directional antennas working at
2-8 GHz are used at both Tx and Rx sides. First, the M8190A generates pseudo noise (PN) sequences as 
channel probing signals with chip rate of 320 Mbps. 
The intermediate frequency (IF) signal is obtained by mixing received radio frequency signal with the local oscillator signal provided by N5173B, amplified by the IFA, 
and then sampled by M9703B in real-time at the rate of 1.6 GSa/s. Finally, the down-converter operation is completed by Keysight 89600 software. 
The measurement system is synchronized by the 1 pulse per second (PPS) produced by the high GPS Rubidium clock at Tx and Rx sides.\par

\begin{table}[!t]
\caption{CONFIGURATION AND PARAMETERS OF CHANMEL MEASURMENT}
\begin{center}
\begin{tabular}{|c|c|}
\hline
\textbf{Configuration} & \textbf{Description} \\
\hline
Center frequency (GHz) & 2.4, 5, 6 \\
\hline
Bandwidth (MHz) & 320 \\
\hline
Delay resolution (ns) & 3.125 \\
\hline
PN code length (chips) & 4800 \\
\hline
Maximum output power (dBm) & 20 \\
\hline
Tx height (m) & 1.95 \\
\hline
Rx height (m) & 1.45 \\
\hline
Number of antenna & Tx: 1, Rx: 1\\
\hline
Antenna gain (dBi) & 2 \\
\hline 
\end{tabular}
\label{table1}
\end{center}
\end{table}

\subsection{Measurment Environment}
\begin{figure}[htbp]
    \centerline{\includegraphics[width=8cm]{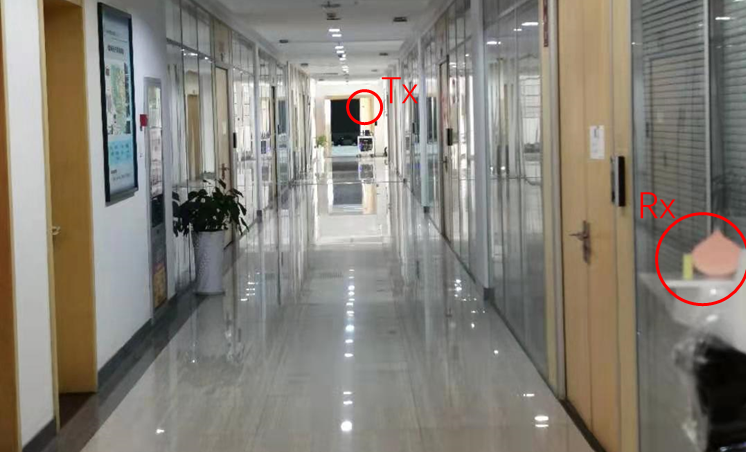}}
    \caption{Multi-frequency wideband indoor corridor channel measurements environment}
    \label{environment}
\end{figure}

\begin{figure}
    \centering
    \centerline{\includegraphics[width=6cm,height=10cm]{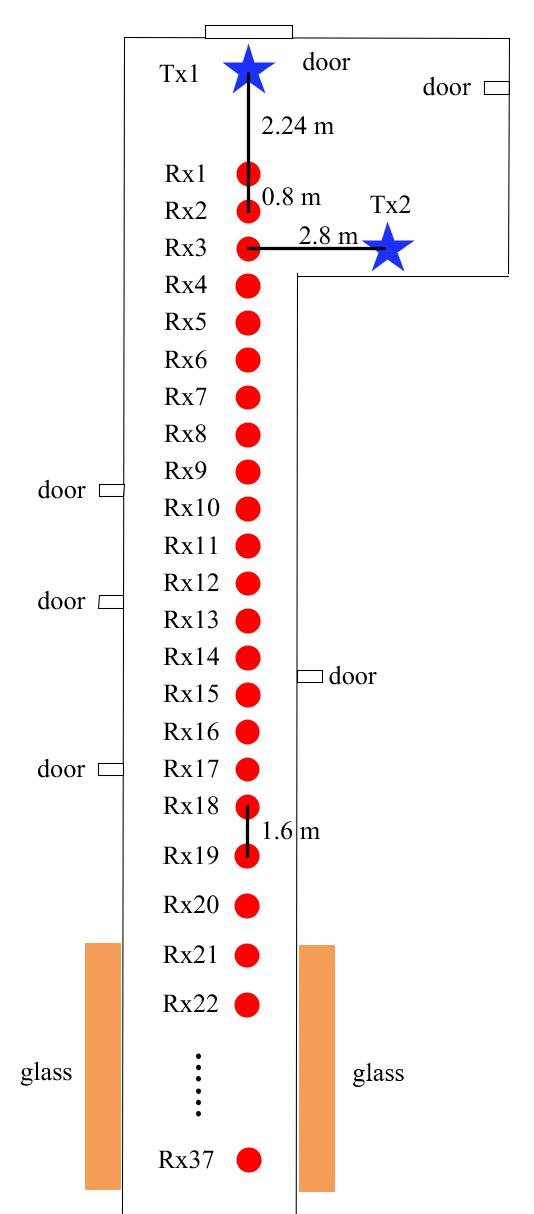}}
    \caption{The layout of the corridor and the locations of the Tx and Rx}
    \label{layout}
\end{figure}

This channel measurement campaign was carried out in a typical indoor corridor scenario with 41 m long, as shown in
Fig. \ref{environment}. Both Tx and Rx antennas were placed on the trolley, and the height of Tx and Rx antennas are 1.95 m and 1.45 m, respectively. 
The corridor was flanked by wooden doors, glass walls, and concrete walls. The layout of this scenario and the locations of Tx and Rx are illustrated in Fig. \ref{layout}. 
Tx1 and Tx2 represent the position of Tx in LOS and NLOS scenarios, respectively, and the red dots represent the position of Rx. 
In the LOS scenario measurement, Rx was moved from position 1 to 37, from position 1 to 21 with an interval of 0.8 m between each two positions, and from position 21 to 37 with an interval of 1.6 m between each two positions. 
While in the NLOS scenario measurement, Rx was moved from position 3 to position 37. 
We measured each position 5 times, and each time with 400 snapshots. 
During this measurement, both Tx and Rx remain stationary.

\section{MEASURMENT DATA PROCESSING METHODS}
\subsection{Data Post-Processing}
Before the actual measurement, the channel sounder needs to be calibrated by back-to-back connection to remove the response of the measurement system.
For convenience, the transmitted signal is denoted by $u(t)$, the back-to-back system response is represented by $g_{th}(t)$, and
the channel impulse response (CIR) is denoted by $h(t)$. The method of data post-processing is the same as \cite{a10}. The received signal obtained by direct connection calibration of system is  
\begin{equation}
    y_{th}(t) = u(t) * g_{th}(t).
\end{equation}
where * represents the convolution operation. The wireless received signal is 
\begin{equation}
    y_{rx}(t) = u(t) * g_{th}(t) * h(t).
\end{equation}
\indent The frequency domain response are
\begin{equation}
    Y_{th}(f) = U(f)G_{th}(f)
\end{equation}
\begin{equation}
    Y_{rx}(f) = U(f)G_{th}(f)H(f).
\end{equation}
\indent Then, the CIR is obtained by inverse fast Fourier transform (IFFT) as
\begin{equation}
    h(t) = \text{IFFT}\{H(f)\} = \text{IFFT}\{Y_{rx}(f)/Y_{th}(f)\}.
\end{equation}
\subsection{PL model}
The closed-in (CI) free space reference PL model, FI PL model and free space PL model are three commonly used PL models. 
The CI PL model is 
\begin{equation}\label{CI}
    PL(d,f) (dB) = 32.4 + 20 \text{log}_{10}(f) + 10n \text{log}_{10}(d) + X_\sigma.
\end{equation}
where $f$ is the center frequency, $n$ is the PLE, $d$ is the distance between Tx and Rx , and $X_\sigma$ is the shadowing fading (SF), which 
can be modeled by a normal distribution of $N(0,\sigma)$, $\sigma$  represents the standard variance of SF (dB).\par
The FI PL model is given by 
\begin{equation}\label{FI}
    PL(d) (dB) = 10\alpha \text{log}_{10}(d) + \beta + X_\sigma.
\end{equation}
where $\alpha$ represents the PLE and $\beta$ represents the path loss offset determined from the measured data.\par
The free space path loss model is 
\begin{equation}
    PL(d,f) (dB) = 32.4 + 20\text{log}_{10}(f) + 20\text{log}_{10}(d).
\end{equation}\par

\subsection{Channel statistical characteristics}
In order to reduce the impact of noise, CIR with signal to noise ratio (SNR) less than 25 dB are eliminated, the remaining CIRs of 5 measurements at each position are averaged to obtain the APDP, which is calculated by 
\begin{equation}
    APDP = \frac{1}{N}\sum_{n=1}^{N}|h_n(\tau)|^2.
\end{equation}
where $h_n(\tau)$ represents the averaged CIR of the $n$th measurement, $N$ is the total number of measurement times, here is 5, and $| \cdot |$ stands for absolute value.\par
The  peak  search  algorithm  is  used  to  extract  multipath components (MPCs) based on the APDP which has eliminated the effect of small scale fading. 
The larger value of noise floor plus 6 dB [13] and maximum power minus 25 dB is selected as the multipath power decision threshold. 
The received power is  calculated  by  summing  the  power  of  all  MPCs 
\begin{equation}
    P_r = \sum_{l=1}^{L}P_l.
\end{equation}
where $P_{l}$ represents the power of the $l\text{th}$ path and $L$ represents the total number of MPCs. Then  the PL is calculated by
\begin{equation}
    PL(dB) = -P_r + P_t + G_t + G_r + P_{thr} - P_{tht} + G_{att}.
\end{equation}
where $P_r$ and $P_{thr}$ are the received power of channel measurement and system calibration, respectively.
$P_t$ and $P_{tht}$ represent the transmitted power of channel measurement and system calibration, respectively.
$G_r$ and $G_t$ are the gains of Tx and Rx antennas, respectively, both of them are 2 dBi, 
and $G_{att}$ represents the attenuation value of attenuator during system calibration. \par

The root mean square (RMS) DS is defined as
\begin{equation}
    \tau_{DS} = \sqrt{\frac{\sum_{l = 1}^{L}P_{l}(\tau_{l}-\overline{\tau})^2}{\sum_{l = 1}^{L}P_{l}}}. 
\end{equation}
\begin{equation}
    \overline{\tau} = \frac{\sum_{l = 1}^{L}P_{l}\tau_{l}}{\sum_{l = 1}^{L}P_{l}}.
\end{equation}
where  $\tau_{l}$ represents the delay of the $l\text{th}$ paths.\par

The frequency domain moment estimation method of broadband channel proposed in \cite{a13} is used to estimate KF, which is suitable for the KF estimation in the static case. 
The frequency domain moment estimation method is calculated by
\begin{equation}
    G_a = \frac{1}{I}\sum_{i = 1}^{I}\left\lvert H_{i} \right\rvert^2.  
\end{equation}
\begin{equation}
    G_v = \frac{1}{I-1}\left(\sum_{i = 1}^{I} \left\lvert H_{i}\right\rvert^4 - iG_{a}^2 \right). 
\end{equation}
\begin{equation}
    \widehat{K} = \frac{\sqrt{G_{a}^2 - G_{v}}}{G_{a}-\sqrt{G_{a}^2-G_{v}}}.  
\end{equation}
where $H_{i}$ is the response of the $i$-th frequency point, $I$ is the number of narrowband spectral samples, $G_{a}$ and $G_{v}$ represent the
first and second moments of the spectral samples respectively, and $\widehat{K}$ represents the estimated KF.\par

\section{RESULTS AND ANALYSIS}
\subsection{APDP}
\begin{figure*}[htbp]
\centering
\subfigure[]{
\includegraphics[width=5cm]{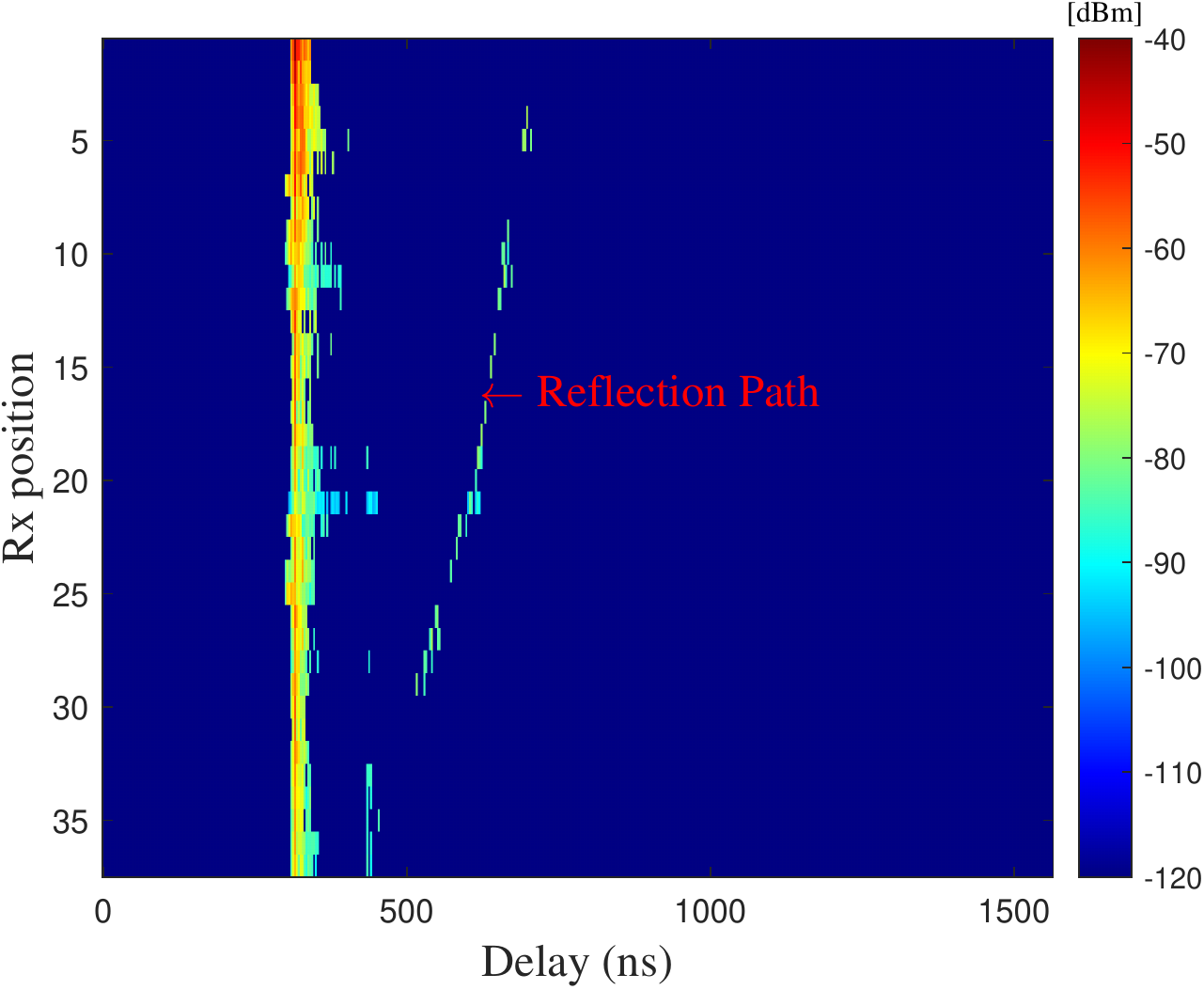}
}    
\quad
\subfigure[]{
\includegraphics[width=5cm]{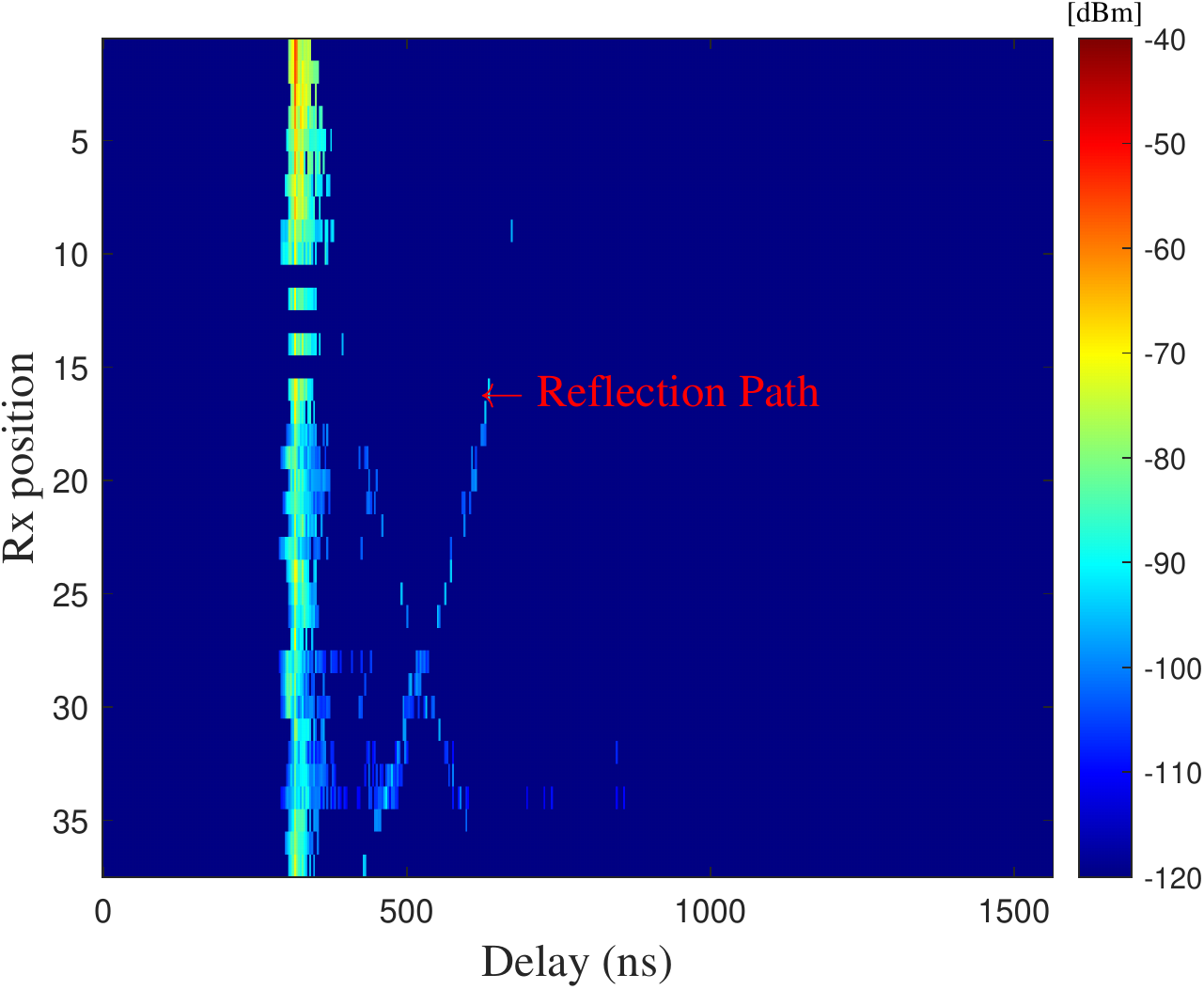}
}
\quad
\subfigure[]{
\includegraphics[width=5cm]{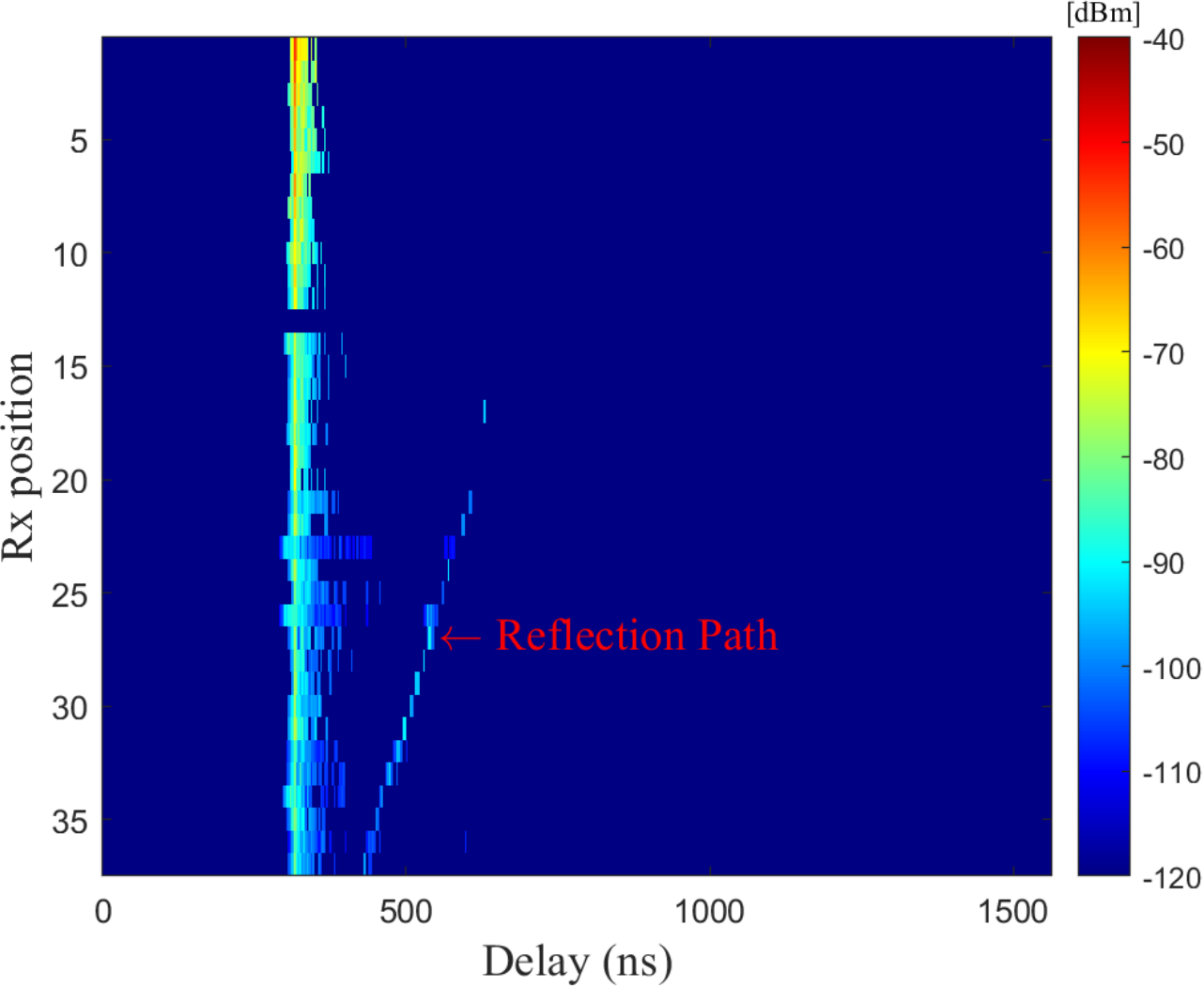}
}
\caption{APDP at (a) 2.4 GHz, (b) 5 GHz, and (c) 6 GHz.}
\label{APDP}
\end{figure*}

The results of APDPs in LOS scenario at three frequency bands are shown in Fig. \ref{APDP}.  
The delay of LOS path at all measurement positions is shifted to the same value, 
and the data points above 6 dB above noise floor are detected to ensure the image can show the effective multipath more clearly.\par 
It can be seen from Fig. \ref{APDP} that APDPs of the three frequency bands have obvious reflection path, and the delay of this reflection path decreases with the increase of distance between Tx and Rx, indicating that the path comes from the end of corridor. 
The reflection paths of 2.4, 5 and 6 GHz are first seen at positions 4, 9 and 17, respectively, and the distance of reflection path first seen is getting shorter and shorter, which means that the attenuation of reflection path increases with increasing frequency.
In addition, it can be seen that the power of LOS path decreases with increasing the distance from Tx between Rx.
  
\subsection{PL}
\begin{figure*}[htbp]
\centering
\subfigure[]{
\includegraphics[width=5cm]{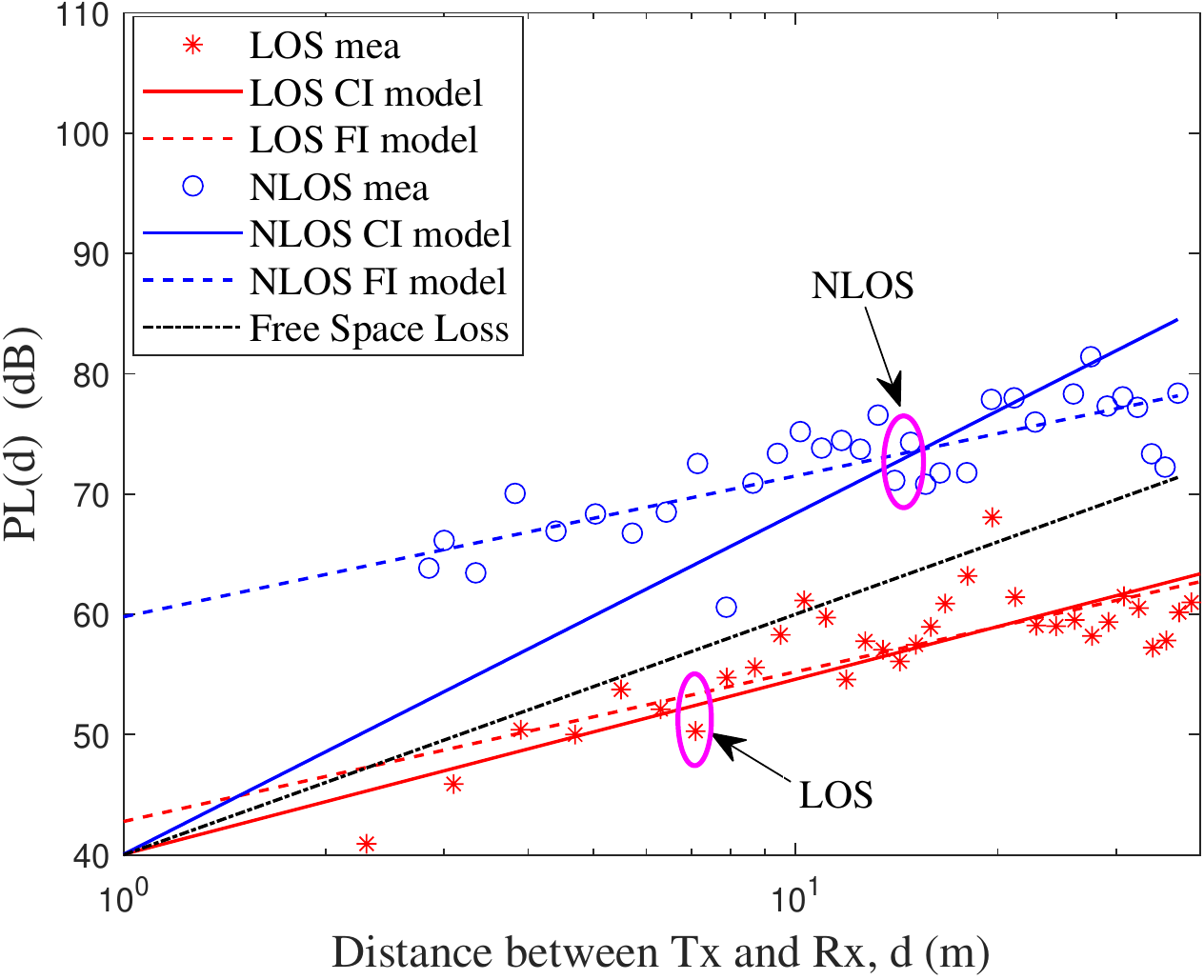}
}
\quad
\subfigure[]{
\includegraphics[width=5cm]{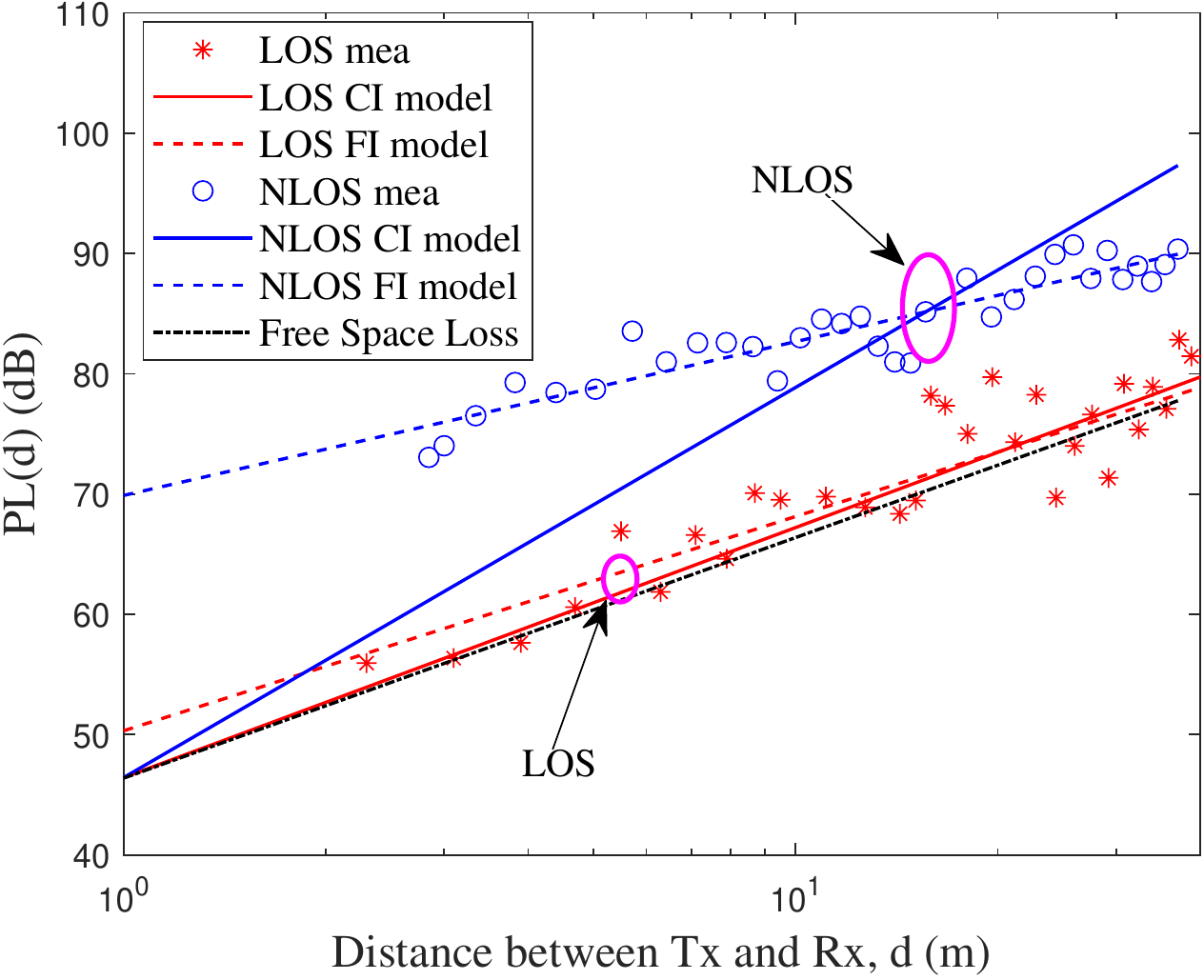}
}
\quad
\subfigure[]{
\includegraphics[width=5cm]{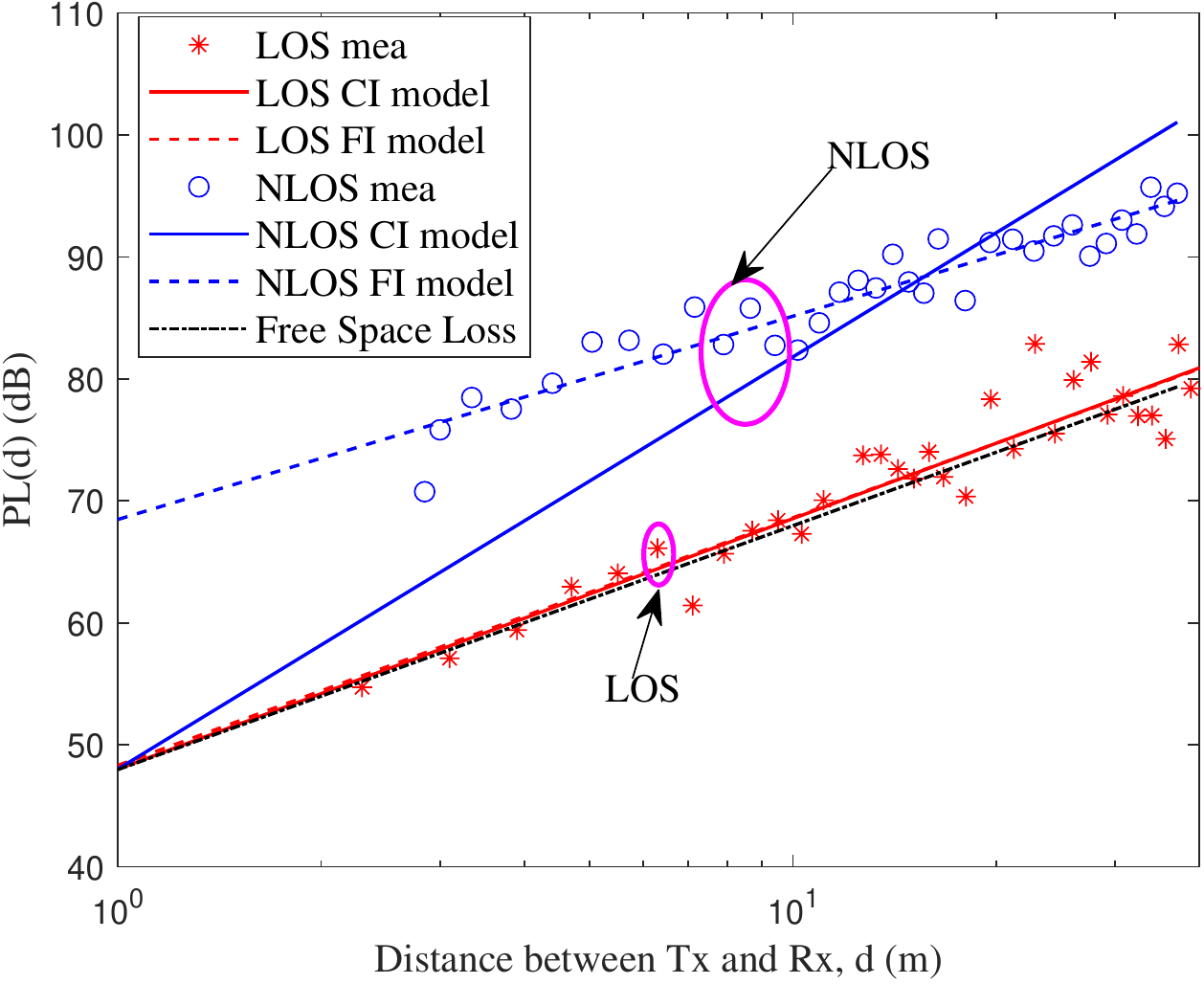}
}
\caption{PL measurment and fitting results at (a) 2.4 GHz, (b) 5 GHz, and (c) 6 GHz.}
\label{Pathloss}
\end{figure*}

The measurement results of PL at 2.4, 5, and 6 GHz frequency bands in LOS and NLOS scenarios are shown in Fig. \ref{Pathloss}. 
The detailed parameters of CI and FI models are listed in Table \ref{table2}. \par 
It can be seen from Fig. \ref{Pathloss} that the
prediction accuracy of FI model in LOS scenario is similar to CI model, however it is better in NLOS scenario. In NLOS scenario,
the CI model will underestimate the PL when the Tx and Rx are close to each other and the CI model will overestimate the PL when the distance
between Tx and Rx is far.\par
In addition, the PL at 2.4 GHz is smaller than free space PL in LOS scenario, however the PL of 5 and 6 GHz is basically the same as free space PL, this may be caused by the stronger waveguide effect at 2.4 GHz.
The PL in NLOS scenario is larger than that in LOS scenario. Both in LOS and NLOS scenarios, the PL at 2.4 GHz is
the smallest, and the PL at 5 and 6 GHz is close. By observing Table \ref{table2}, it can be found that the PLE and $\beta$ value in the FI model 
tend to increase with the increase of frequency, which is consistent with \cite{MF1}. 

\begin{table}[!t]
    \caption{CI MODEL AND FI MODEL FOR 2.4, 5 AND 6 GHz BANDS}
    \begin{center}
    \begin{tabular}{|c|c|c|c|c|}
    \hline
    \multicolumn{2}{|c|}{\textbf{Frequency band (GHz)}}&\textbf{2.4} & \textbf{5} & \textbf{6}\\
    \hline
    \textbf{FI model}& \textbf{LOS} & 1.25, 41.23,  & 1.75, 48.93, & 2.03, 48.33,  \\
    \textbf{parameters,} &  & 2.96 & 3.18 & 2.39\\
    \cline{2-5}
    \textbf{($\alpha$,$\beta$,$\sigma$)}& \textbf{NLOS} & 1.17, 57.46, & 1.27, 67.22,  & 1.66, 67.91,  \\
    &  &  2.99 &  1.85 & 1.87 \\
    \hline
    \textbf{CI model }&\textbf{LOS} & 1.45, 3.05 & 2.08, 3.59 & 2.05, 2.39 \\
    \textbf{parameters,} & & & & \\
    \cline{2-5}
    \textbf{($n$,$\sigma$)}&\textbf{NLOS} & 2.83, 6.34 & 3.24, 6.91 &3.37, 6.03 \\
    \hline
    \end{tabular}
    \label{table2}
    \end{center}
\end{table}

\subsection{DS}
The RMS DS in LOS and NOLS scenarios is listed in Table \ref{table3}. 
It can be found that the mean values of DS at three frequency bands are close to each other in the LOS scenario, and there is no obvious frequency dependence.
This is in line with the conclusion in \cite{MF1}.\par
Comparing DS in LOS and NLOS scenarios, we can find that the standard deviation (STD) values of DS in NLOS scenario are smaller than that in LOS scenario at three frequency bands, and the mean values of DS in LOS and NLOS scenarios at 2.4 and 5 GHz are basically the same. 
This may be due to the existence of a strong reflection path in the LOS scenario.
The delay of this path is much larger than that of LOS path, and the power is comparable to LOS path, increasing the mean value of DS in LOS scenario.
Moreover, this reflection path does not exist at all positions, which increases the STD value of DS in LOS scenario.
\subsection{KF}
\begin{figure}[!t]
    \centerline{\includegraphics[width=7.5cm]{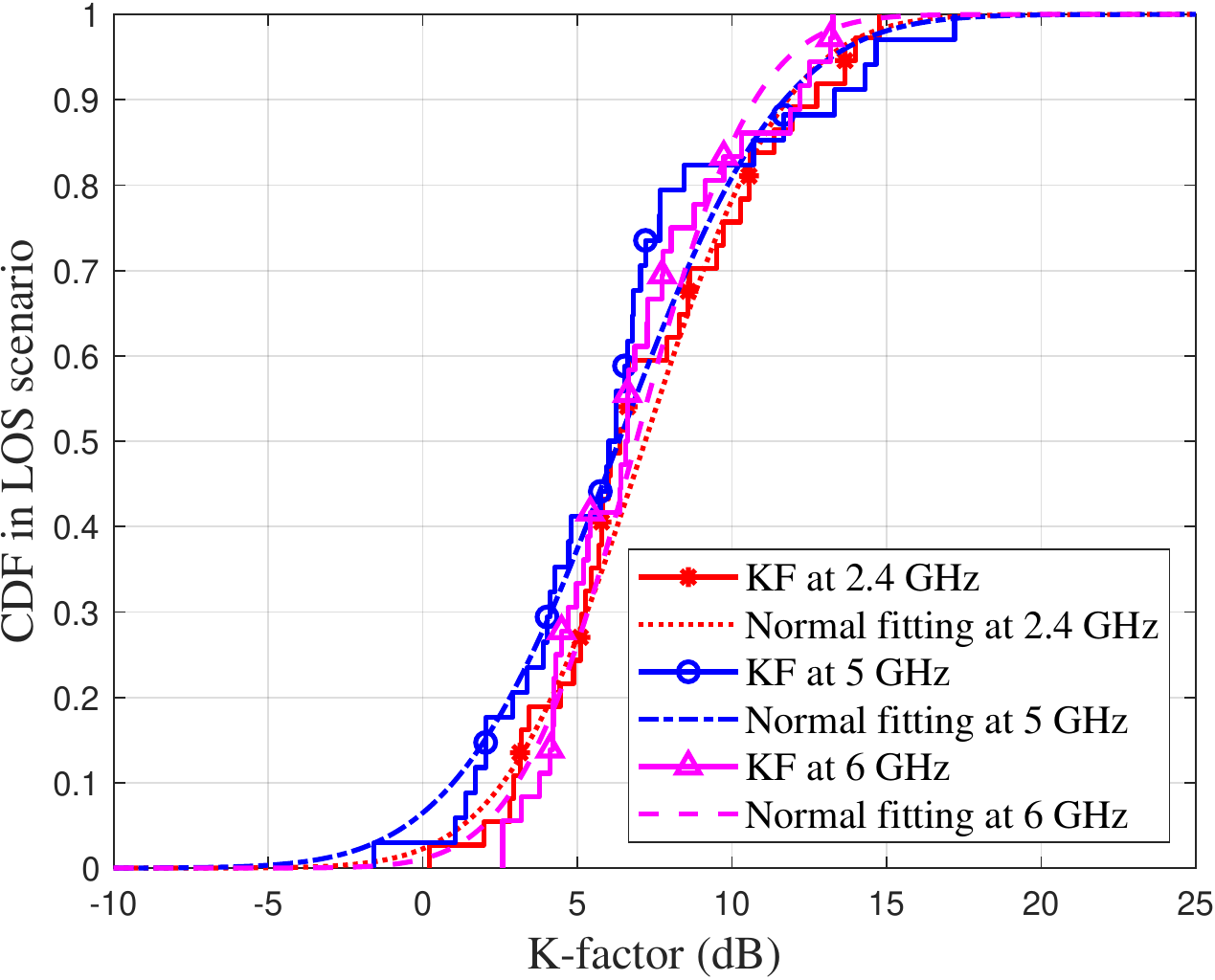}}
    \caption{KF at 2.4, 5, 6 GHz frequency bands in LOS scenario}
    \label{KF}
\end{figure}

Fig. \ref{KF} shows the cumulative distribution function (CDF) of KF at three frequency bands in LOS scenario, which can be fitted well by normal distribution.\par
As can be seen from Fig. \ref{KF} that the KF values of the three frequency bands are very close, which is consistent with the changing trend of DS.
This may be because the environment does not change significantly during the channel measurement. 
A summary of detailed parameters description is listed in table~\ref{table3}. 
\begin{table}[htbp]
    \caption{PARAMETERS ANALYSIS RESULTS FOR 2.4, 5, AND 6 GHz BANDS IN LOS SCENARIO}
    \begin{center}
        \begin{tabular}{|c|c|c|c|c|c|}
            \hline
            \multicolumn{3}{|c|}{\textbf{Frequency (GHz)}}&\textbf{2.4}&\textbf{5}&\textbf{6}\\
            \hline
            \multirow{4}{*}{\textbf{DS (ns)}} & \multirow{2}{*}{\textbf{LOS}} & \textbf{Mean}&19.67 & 20.48 & 18.75\\
            \cline{3-6}
            &  &\textbf{STD}& 14.89& 13.20 & 14.17 \\
            \cline{2-6}
            & \multirow{2}{*}{\textbf{NLOS}} & \textbf{Mean} & 20.83 & 20.64& 28.60 \\
            \cline{3-6}
            & & \textbf{STD} & 7.85 & 7.36 & 9.52\\ 
            \hline
            \multirow{2}{*}{\textbf{KF (dB)}} & \multirow{2}{*}{\textbf{LOS}} & \textbf{Mean}& 7.19 & 6.34 & 6.90\\
            \cline{3-6}
            &  &\textbf{STD} & 3.59 & 4.18 & 3.00 \\
            \hline
            \end{tabular}
            \label{table3}    
    \end{center}
\end{table}
\section{CONCLUSION}
Channel measurements at 2.4, 5, and 6 GHz in indoor corridor scenarios are conducted in this paper. Channel characteristics such as APDP, PL, DS, and KF at three frequency bands were analyzed. A distinct reflection path 
can be seen in the APDP of three frequency bands in the LOS scenario. In addition, it can be observed that the PLE and PL offset in the FI PL model have a tendency to increase
with frequency. In the NLOS scenario, the fitting accuracy of FI PL model is better than that of CI PL model, because the CI PL model will underestimate the value 
of PL when the distance between Tx and Rx is close, and overestimate the PL value when the Tx is far away from Rx.
However, no significant frequency dependence was observed for DS and KF.\par

\section*{ACKNOWLEDGEMENT}

\small{This work was supported by the National Key R\&D Program of China under Grant 2018YFB1801101, the National Natural Science Foundation of China (NSFC) under Grant 61960206006 and
 Grant 61901109, the Frontiers Science Center for Mobile Information Communication and Security, the High Level Innovation and Entrepreneurial Research Team Program in Jiangsu, the High
 Level Innovation and Entrepreneurial Talent Introduction Program in Jiangsu, the Research Fund of National Mobile Communications Research Laboratory, Southeast University, under Grant 2020B01, the
  Fundamental Research Funds for the Central Universities under Grant 2242021R30001, and the EU H2020 RISE TESTBED2 project under Grant 872172, 
  the National Postdoctoral Program for InnovativeTalents under Grant BX20180062.}

\end{document}